 \documentclass[final,3p,times,twocolumn,sort&compress]{elsarticle}

 \usepackage{graphicx,amssymb,amsmath,amsthm,lineno}
 \usepackage{dcolumn,multirow}
 \usepackage{subfigure}
 \usepackage{threeparttable}

\biboptions{square}

\journal{Physics Letters A}

\begin{document}

\begin{frontmatter}



\title{Influence of the Long-Range Forces in Non-Gaussian Random-Packing Dynamics}


\author[ufsa]{Carlos Handrey Araujo Ferraz\corref{cor1}}
\ead{handrey@ufersa.edu.br}
\cortext[cor1]{Corresponding author}
\address[ufsa]{Exact and Natural Sciences Center, Universidade Federal Rural do Semi-\'Arido-UFERSA, PO Box 0137, CEP 59625-900, Mossor\'o, RN, Brazil}

\begin{abstract}
 In this paper, we perform molecular dynamics (MD) simulations to study the random packing of spheres with different particle size distributions. In particular, we deal with non-Gaussian distributions by means of the L{\'e}vy distributions. The initial positions as well as the radii of five thousand non-overlapping particles are assigned inside a confining rectangular box. After that, the system is allowed to settle under gravity towards the bottom of the box. Both the translational and rotational movements of each particle are considered in the simulations. In order to deal with interacting particles, we take into account both the contact and long-range cohesive forces. The normal viscoelastic force is calculated according to the nonlinear Hertz model, whereas the tangential force is calculated through an accurate nonlinear-spring model. Assuming a molecular approach, we account for the long-range cohesive forces using a Lennard-Jones(LJ)-like potential. The packing processes are studied assuming different long-range interaction strengths. 
\end{abstract}


\end{frontmatter}


\section{Introduction \label{sec:int}}

Over the years, scientific research has revealed that systems such as molecular liquids, colloids, and granular media possess, in certain conditions, a similar phenomenological behavior with respect to their glassy phase transitions. Colloidal systems composed of hard spheres display a fluid-like phase with density $\phi$ from $0$ to intermediate values, a freezing crystallization at $\phi\simeq 0.494$, and a melting transition at $\phi\simeq 0.545$~\cite{pusey1986}. Above this melting point, the colloidal system can be compressed until the close-packing point reaches $\phi\simeq 0.74$, where the equilibrium state is crystalline. Remarkably, a small amount of polydispersity (i.e., particles with slightly different sizes) in the system can effectively prevent crystallization~\cite{pusey2009}. As a consequence, the system can be easily ``super-compressed" above the freezing transition without nucleation or crystal growth.  It has also been observed that such systems exhibit relaxation time scales that increase rapidly with increasing $\phi$. Likewise, a polydispersive granular fluid, at high packing density, displays a relaxation (or diffusion) time that rises rapidly with increasing density $\phi$, without an evident change in its structural properties. This universal characteristic has been referred to as the jamming state~\cite{liu1998}. Typically for colloids and granular media~\cite{Bideau2004}, a ``jammed" phase could be obtained either by increasing the packing density or by decreasing the external drive (e.g., shearing and tapping)

The random packing of spherical particle, in particular, has been an interesting tool used to capture the underlying behavior of more complex phenomena for applications in physics and materials engineering such as modeling ideal liquids~\cite{bernal60,bernal64}, amorphous materials~\cite{finney76,angell81}, granular media~\cite{herrmann95}, emulsions~\cite{pal2008}, glasses~\cite{lois2009}, jamming~\cite{hern2003}, living cells~\cite{torquato2000}, ceramic compounds~\cite{mcgeary61, hamad90} and sintering processes~\cite{helle85,nair87}. Understanding the structure of random close-packed particles is important because its physical properties may depend on the packing features such as packing density and porosity. The packing density (i.e., the volume ratio occupied by particles to the total aggregate) is affected by the particle size distribution, particle shape and long-range cohesive forces. In general, random packing structures possess packing densities that increase with increasing width of the size distribution~\cite{sohn1968,schaertl1994,hermes_2010,santos2014}, increasing sphericity, and decreasing long-range cohesive forces. For micro-sized particles, or smaller, both van der Waals and electrostatic forces play an important role in particle rearrangements as they dominate the dynamical packing process~\cite{visser89,israelachvili92}, forming local particle clusters~\cite{yen91, yang2000, cheng2000, jia12} that can eventuate into large percolation clusters~\cite{handrey2018} depending on the nature of the particles involved.

There have been few prior experimental and computational studies concerning the micro-sized particles packing in which long-range cohesive forces have to be taken into account to describe the adequate behavior of the colliding particles involved in these dynamical processes. Forsyth {\it et al}~\cite{forsyth2001} experimentally investigated the influence of van der Waals forces in hard-sphere packing; however, they did not take into account neither the impacts caused by electrostatic force nor polydispersity. Liu {\it et al}~\cite{liu1999} performed computational simulations to address the centripetal packing of mono-sized spheres. Yen and Chaki~\cite{yen91}, Cheng {\it et al}~\cite{cheng2000} and Yang {\it et al}~\cite{yang2000} each applied a simplified version of the so-called distinct element method~\cite{cundall79} to study the effects of both van der Waals and frictional forces present in hard-sphere packing processes but also did not consider particle size distributions in their investigations. More recently, a computational study~\cite{jia12} has considered particle packing dynamics using Gaussian size distribution where the van der Waals forces were calculated using the standard Hamaker form~\cite{hamaker1937} without the inclusion of the electrostatic forces between particle pairs or non-Gaussian effects in particle size distributions. Electrostatic interactions are, however, quite important because of their fundamental role in governing the properties of many systems, including soft matter, colloidal suspensions, electrolyte solutions and various biological systems~\cite{dhont2002soft, gompper2006soft}.

\begin{figure}[!t]
 \centering
 \includegraphics[scale=0.28]{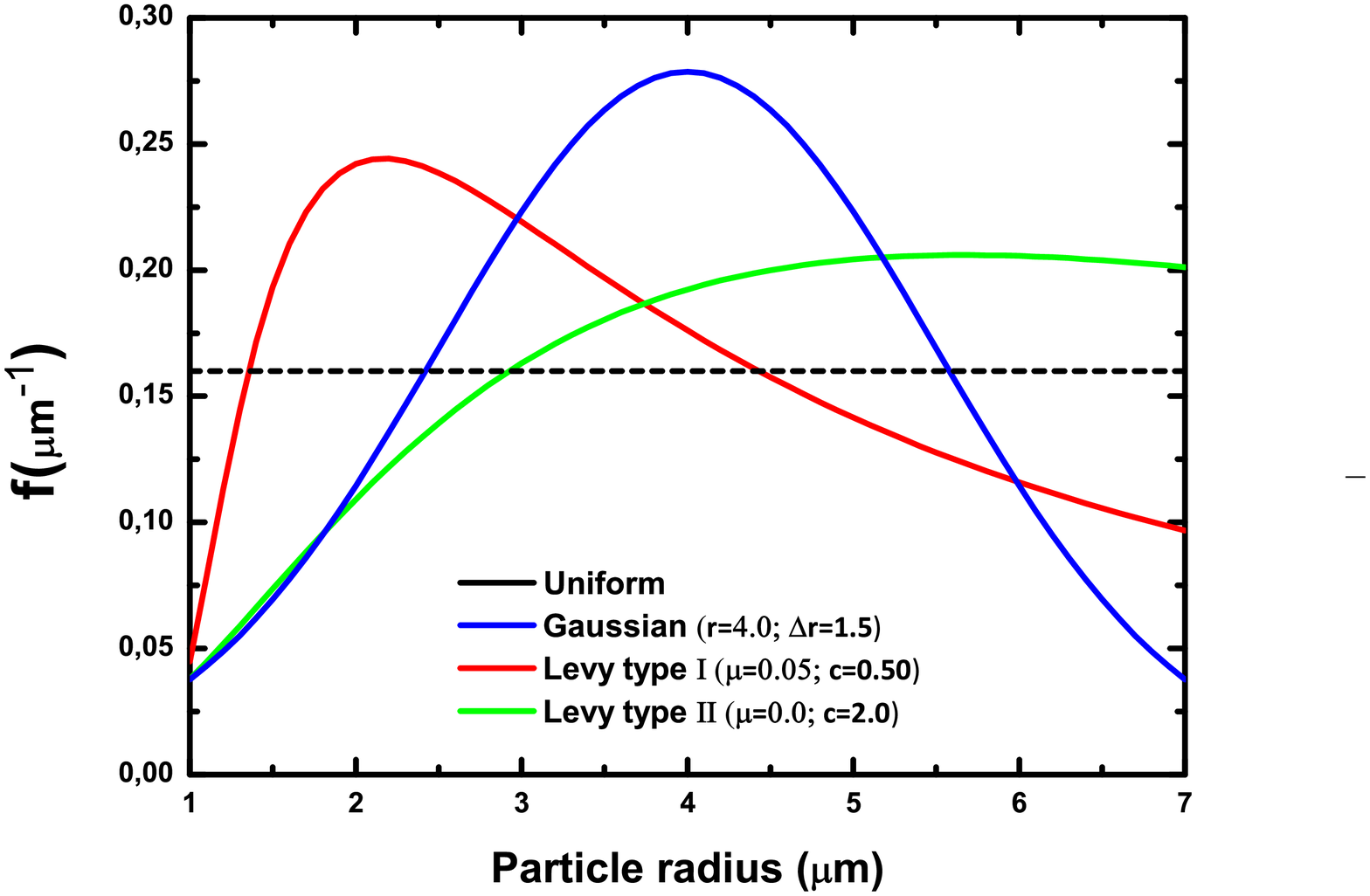}
 \caption{ Probability distributions applied in this study. L{\'e}vy type I (red line), L{\'e}vy type II (green line), Gaussian (blue line) and uniform distribution (black dashed line). All distributions have been normalized within the $r$ value range of interest.} \label{fig:01}
\end{figure}

Alternatively, one can represent the long-range cohesive forces present in many molecular systems by coarse-grained intermolecular potentials, notably by avoiding the full atomic representation of molecules or macromolecules, to find a description for the interaction at either long or coarse length scales. This approach, despite its simplicity, has been successfully used to model systems such as liquid crystal~\cite{zannoni2001}, proteins~\cite{monticelli2008} and water molecules~\cite{lobanova2015}. In most of these studies, a modified version of the Lennard-Jones (LJ) potential has been employed. Hence, one begins to wonder whether such an approach could also be applied to modeling micro-sized particles and their long-range interactions. This ``molecular" approach, particularly for large particulate systems, can be guaranteed as long as we realize that when two microspheres (with radius $R$) are separated by a certain distance $D >> R$, the effective potential ($\Phi$) is analogous to that between two molecules, i.e., falling off as $\Phi(D)\varpropto -1/D^6$~\cite{israelachvili92, lyklema91}. Assuming the validity of this modified LJ approximation, we therefore are able to account for the long-range forces involved in the packing process. This approximation will allow us to study a variety of different packing cases by considering LJ particles with different potential well depths, which play a dominant role in the strength of these long-range forces.

In this paper, we perform molecular dynamics (MD) simulations to study the packing process of spheres using different particle size distributions. Here, while we deal with non-Gaussian distributions through the rescaled L{\'e}vy profiles and uniform distributions, we also account for particle packing utilizing Gaussian distributions in order to compare the different packing features.  Both the translational and rotational movements of each particle are also considered in the simulations. In order to deal with interacting particles, we take into account both the contact and long-range cohesive forces. The contact forces result from the deformation of the colliding particles, which can be decomposed into two main types: the normal viscoelastic force and the tangential force. The normal viscoelastic force is calculated according to the nonlinear Hertz model~\cite{popov2010, brilliantov1996}, whereas the tangential force is calculated through a nonlinear-spring model that is derived from the Mindlin--Deresiewicz theory~\cite{mindlin1953}.   By assuming a molecular approach, we account for the long-range forces using a modified LJ potential. The packing processes were studied by applying different long-range interaction strengths. It is worthy mentioning that more sophisticate approaches such as the Johnson-Kendall-Roberts (JKR) model~\cite{johnson1985} has been used for treating interacting particles whereby contact adhesive effects are also considered. For simplicity, these effects were not taken into account in our simulations.

We performed statistical calculations of the different quantities studied including packing density, mean coordination number, kinetic energy and radial distribution function (RDF) as the system evolved over time. 

The content of the manuscript is organized as follows. In section~\ref{sec:mms}, we describe in detail, the model and MD simulations. In section \ref{sec:r}, we  present and discuss the results. Finally, in section \ref{sec:c}, we draw the conclusions and give some perspective on possible future developments.

\section{ Model and Molecular Dynamics Simulation \label{sec:mms}}

\begin{figure*}[!t]
\centering
\begin{minipage}[t]{1.0\linewidth}
\centering
\subfigure[]{\label{fig:02a}\includegraphics[scale=0.42, angle=0]{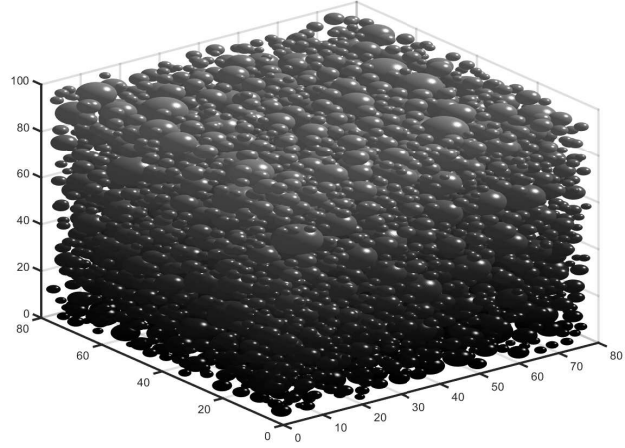}}
\subfigure[]{\label{fig:02d}\includegraphics[scale=0.46, angle=0]{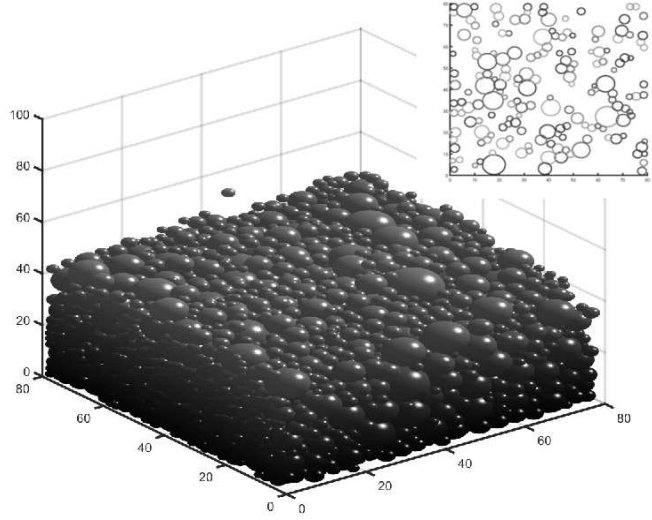}}
\end{minipage}
\caption{Snapshots of a typical packing process with Gaussian distribution inside a $(80 \times 80 \times 100) \, \mu m$ box at the instants $t=0.0 \, ms$ (a) and $t=10.0 \, ms$ (b). The inset of figure~\ref{fig:02d} shows the cross section at $z=20$ (vertical axis) of the formed structure. The parameters used in this simulation are given in Table \ref{table:01} with an interaction strength of $\varepsilon=10.0\, \mu J$.}\label{fig:02}
\end{figure*}

\begin{figure*}[!t]
\begin{minipage} [t]{0.49\linewidth}
\includegraphics*[scale=0.30]{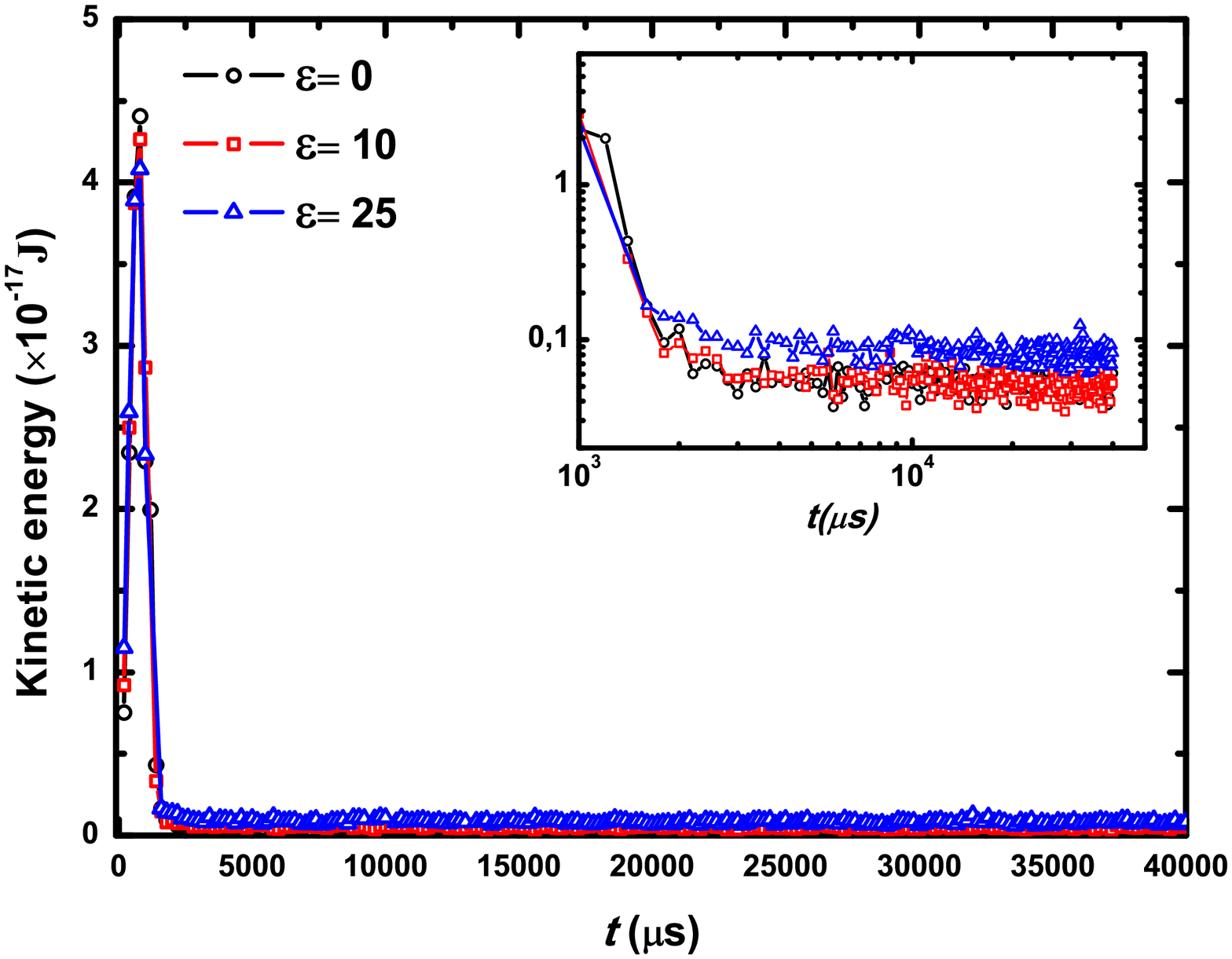}
 \caption{Plot of the average kinetic energy per particle for the Gaussian size distribution as a function of time with four different $\varepsilon$ values. The inset shows the log-log plot of the tail of the curve after $1.0 \, ms$. The noise observed at large times is due mainly to persistent action of the long-range forces on the particles.} \label{fig:04}
\end{minipage}\hfill
\begin{minipage}[t]{0.49\linewidth}
\includegraphics[scale=0.30]{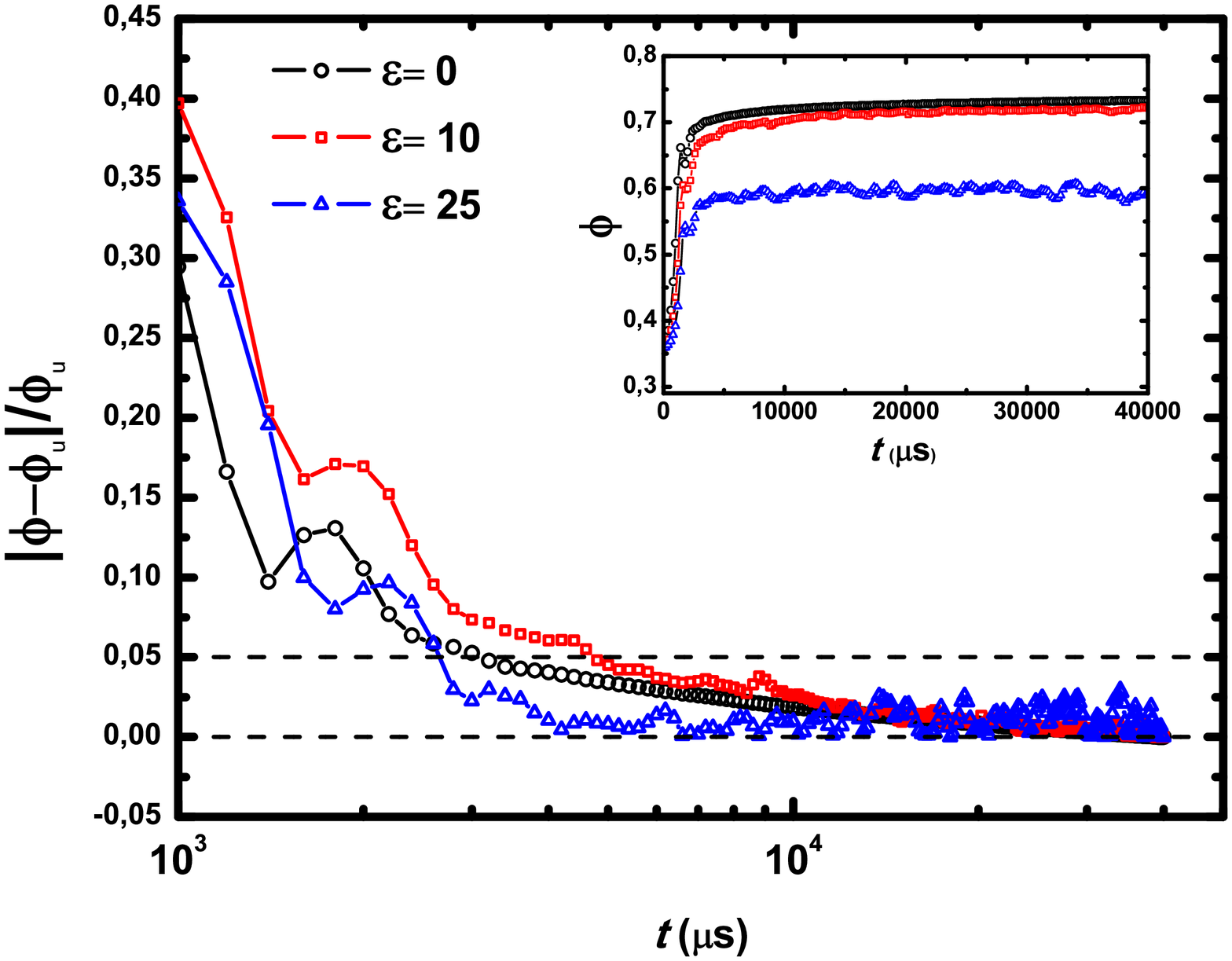}
 \caption{Time evolution of the quantity $|\phi-\phi_{u}|/\phi_{u}$ in the Gaussian case for three different $\varepsilon$ values. The time axis is plotted on a logarithmic scale and the dashed lines help to view the dramatically slow convergence of this quantity. The ultimate packing densities $\phi_{u}$ are obtained at $40 \, ms$. The inset shows the packing density curves on a linear scale. The data points were obtained by taking averages over 10 runs.} \label{fig:44}
\end{minipage}
\end{figure*}

The time evolution of the random packing of spheres was simulated by using the MD method. The equations of motion of an $i$-th particle of mass $m_{i}$ and radius $R_{i}$ are:

\begin{equation} \label{eq:1}
{m_i}\frac{{{d^2}{{\vec r}_i}}}{{d{t^2}}} = \sum\limits_j {(\vec F_{ij}^n + \vec F_{ij}^t + \vec F_{ij}^{LJ}) + {m_i}\vec g }
\end{equation}
and 
\begin{equation} \label{eq:2}
 {I_i}\frac{{d{{\vec \omega }_i}}}{{dt}} = \sum\limits_j {{R_i}\,{{\hat n}_{ij}} \times \vec F_{ij}^t}-\gamma _{r} \,R_i|\vec F_{ij}^n|\,{\vec \omega }_i,
\end{equation}
where $\vec{r}_i$ is the position, $\vec{\omega}_i$ is the angular velocity, $\hat{n}_{ij}$ is the unity vector in the direction $j\rightarrow i$, $\gamma _{r}$ is the rolling friction coefficient and $I_{i}=2/5 \,m_{i}R_{i}^{2}$ is the moment of inertia of the particle. 

In the above equations, $\vec F_{ij}^n$ is the normal viscoelastic force, $\vec F_{ij}^t$ is the tangential friction force, $\vec F_{ij}^{LJ}$ is the LJ force between the $i$- and $j$-th particle, and $\vec g$ is the gravity acceleration. The normal viscoelastic force $\vec F_{ij}^n$ is derived from the nonlinear Hertz theory, and it can written as
\begin{equation} \label{eq:3}
 \vec F_{ij}^n = [\frac{2}{3}E\sqrt {\bar R}\, \delta _{n}^{3/2}-{\gamma _n}E\sqrt {\bar R} \sqrt {{\delta _{n}}} ({{\vec v}_{ij}} \cdot {{\hat n}_{ij}})]{{\hat n}_{ij}},
\end{equation}
where $E$ is the elastic modulus of the two particles, $\bar R=R_{i}R_{j}/(R_{i}+R_{j})$ is the effective radius, $\delta _{n}$ is the deformation which is expressed by
\begin{equation}\label{eq:4}
	\delta _{n}=(R_{i}+R_{j})-(|\vec{r}_{i}(t)-\vec{r}_{j}(t)|),
\end{equation}
 $\vec{v}_{ij}$ is the relative velocity between $i$- and $j$-th particle, and $\gamma _{n}$ is the normal damping  coefficient. 

The tangential friction force $\vec F_{ij}^t$ is calculated according to the Mindlin-Deresiewicz theory as
\begin{equation} \label{eq:5}
\vec F_{ij}^t = {\gamma _t}|\vec F_{ij}^n|\left[ {1 - {{\left( {1 - \frac{{|{\delta _{t}}|}}{{|{\delta _{\max }}|}}} \right)}^{3/2}}} \right]{{\hat t}_{ij}},
\end{equation}
where  $\gamma _t$ is the friction coefficient, ${\hat t}_{ij}$ is the unit vector perpendicular to ${\hat n}_{ij}$, $\delta _{t}$ is the tangential displacement which is determined as
\begin{equation} \label{eq:6}
{\delta _{t}} = \int\limits_{{0}}^{t_{c}} {({{\vec v}_{ij}} \cdot {\kern 1pt} {\kern 1pt} {\kern 1pt} } {{\hat t}_{ij}} + {R_i}\,{{\hat n}_{ij}} \times {{\vec \omega }_i} + {R_j}\,{{\hat n}_{ij}} \times {{\vec \omega }_j})dt,
\end{equation}
where the above integral is calculated during the contact time $t_{c}$ (see below) between the particles. The $\delta _{\max }$ is the maximum tangential displacement and in the condition that $\delta _{t}>|\delta _{\max}|$, the sliding friction takes place between the particles. In Eqs.~\ref{eq:3} and \ref{eq:5}, $E$ and $\delta _{\max }$ are given, respectively, by
\begin{equation}\label{eq:7}
	E=Y/(1-\xi^2)
\end{equation}
and 
\begin{equation} \label{eq:8}
	{\delta _{\max }} = {\gamma _t}\frac{{2 - \xi }}{{2(1 - \xi )}}{\delta _{n}},
\end{equation}
being $Y$ the Young's modulus and $\xi$ the Poisson' ratio. For damped collision~\cite{schwager1998}, the contact time is given by
\begin{equation} \label{eq:9}
	{t_c} = 2.94\;{\Omega ^{ - 2/5}}|{{\vec v}_{ij}}{|^{ - 1/5}}(1 + \frac{1}{{10}}\zeta {\kern 1pt} {\Omega ^{2/5}}|{{\vec v}_{ij}}{|^{1/5}}),
\end{equation}
where $\Omega  = 2/3\,E\,(\bar R/\bar M)$ ($\bar R$ and $\bar M$ are, respectively, the effective radius and mass) and $\zeta=\zeta ({\gamma _n},Y,\xi )$ is a material-dependent constant. For undamped collision~\cite{landau1965}, one just takes $\zeta=0$ in above equation.

\begin{table}[!b]
\centering
\small
\begin{threeparttable}
\caption{\label{table:01} Physical parameters used in the simulations.}
\begin{tabular}{lc}
\hline \hline \\
  Parameter\tnote{a} &  Value \\ \hline \\
  Number of particles ($N$) & $5000$ \\
  Particle size ($R$) & $1.0-7.0\;\mu \,m $ \\
  Particle density ($\rho$) & $2500/\pi \; kg/m^{3}$ \\
	Minimum potential energy ($\varepsilon$) & $0-25.0\;\mu J$\\
  Young's modulus ($Y$) &  $10^{8} \; N/m^{2}$ \\
  Poisson's ratio ($\xi$) & $0.30$ \\
  Normal damping coefficient ($\gamma_{n}$) &  $0.05\; s$ \\
  Tangential damping coefficient ($\gamma_{t}$) &  $0.30$ \\
	Rolling friction coefficient ($\gamma_{r}$) & $0.002$ \\
\hline \hline
\end{tabular}
\begin{tablenotes}
      \item[a]{It is assumed that both particles and walls have the same physical parameters.}
    \end{tablenotes}
\end{threeparttable}
\end{table}

The LJ force between the particles $i$ and $j$ can be evaluated as
\begin{equation} \label{eq:10}
\vec F_{ij}^{LJ} = \frac{{24\varepsilon }}{\sigma }\left[ {2{{\left( {\frac{\sigma }{{{r_{ij}}}}} \right)}^{13}} - {{\left( {\frac{\sigma }{{{r_{ij}}}}} \right)}^7}} \right]{{\hat n}_{ij}},
\end{equation} 
where $r_{ij}$ is the distance between the particles, $\varepsilon$ is the well depth of the LJ potential, which rules the strength of the interaction, and $\sigma=2^{-1/6}(R_{i}+R_{j})$ defines the hard core of the potential. Here, it is important to say that the LJ force is only activated when $r_{ij}>R_{i}+R_{j}$. For $r_{ij}\leq R_{i}+R_{j}$, the contact forces given by Eqs.~\ref{eq:3} and \ref{eq:5} take over control of the particles' driving. It is also worthy mentioning that the continuity and smoothness properties at the transition point between the Eqs.~(\ref{eq:3}) and (\ref{eq:10}) are implicitly assumed since every physical parameter given in Table \ref{table:01}, including the $\varepsilon$ magnitude, has been obtained by trial tests so as to avoid unrealistic behaviors during the particles' interactions. In addition, we have also used a cutoff at $r_{ij}=3\,(R_{i}+R_{j})$ for saving time during the simulations. 

Because of the good accuracy, low computational cost and symplectic feature, a leapfrog scheme~\cite{rapaport95} was used to integrate numerically the Eqs.~(\ref{eq:1}) and (\ref{eq:2}). In order to avoid the complicating effects of the pouring rate, the particles were suspended along the box at the beginning of the simulation. Owing to frictional forces, stable simulations were already achieved by taking a time-step $\delta t = 10^{-6} s$. The average CPU time to update the state of one particle was approximately $0.72 \thinspace \mu s$ on one 3.70 GHz Intel Xeon microprocessor.

\begin{figure*}[!ht]
\centering
\begin{minipage}[t]{1.0\linewidth}
\centering
\subfigure[Gaussian]{\label{fig:03a}\includegraphics[scale=0.29, angle=0]{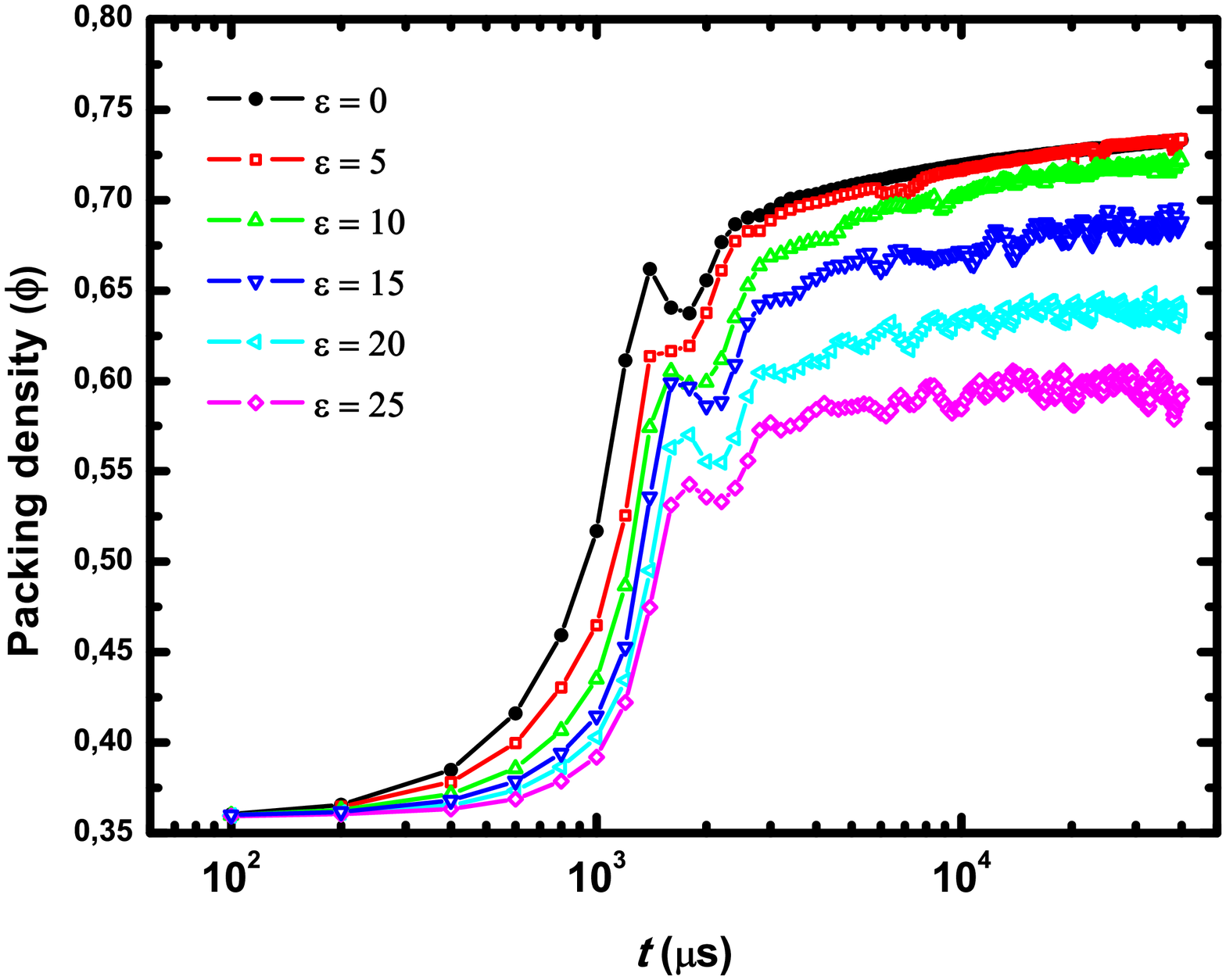}}
\subfigure[L{\'e}vy type I]{\label{fig:03b}\includegraphics[scale=0.29, angle=0]{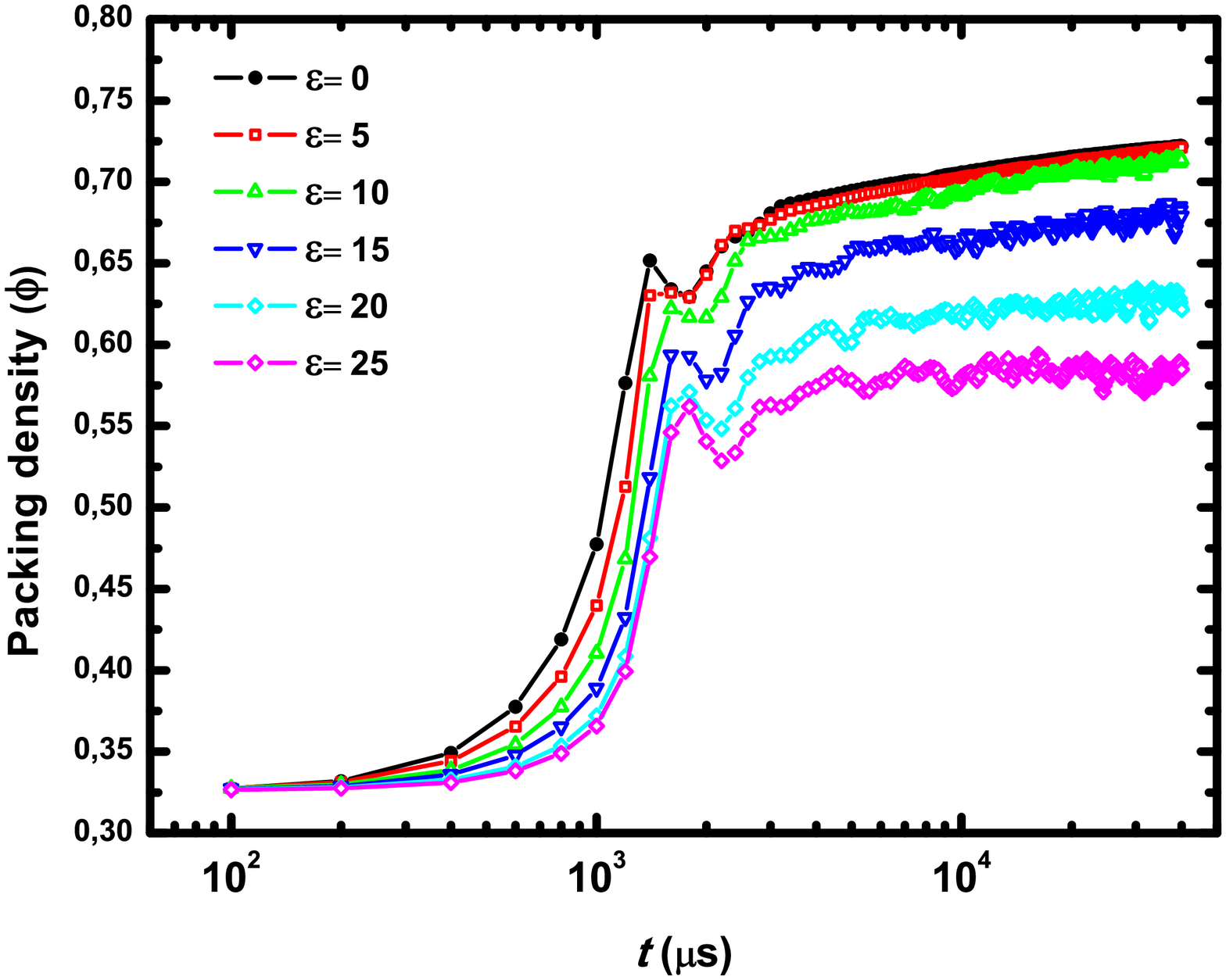}} 
\subfigure[L{\'e}vy type II]{\label{fig:03c}\includegraphics[scale=0.29, angle=0]{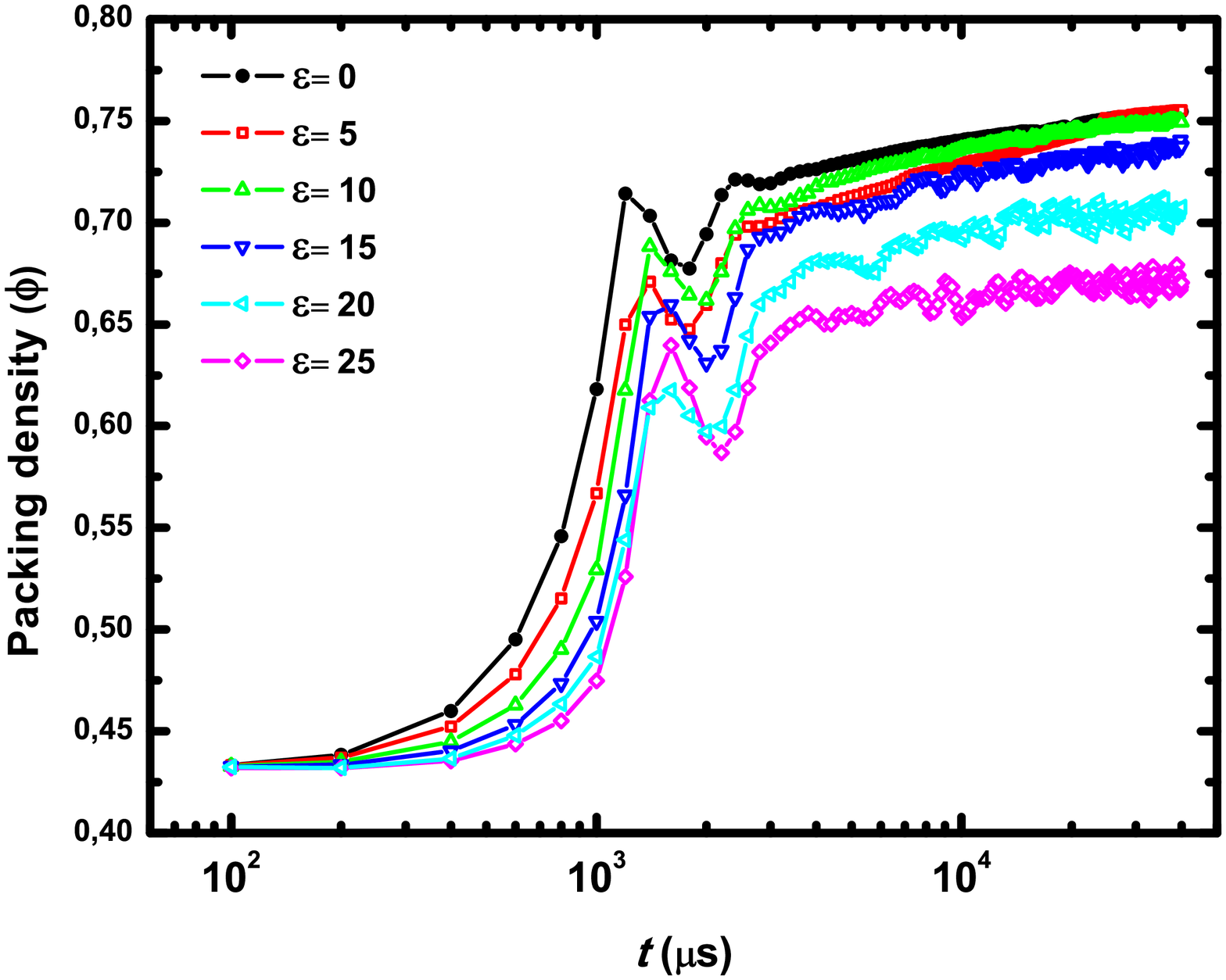}}
\subfigure[Uniform]{\label{fig:03d}\includegraphics[scale=0.29, angle=0]{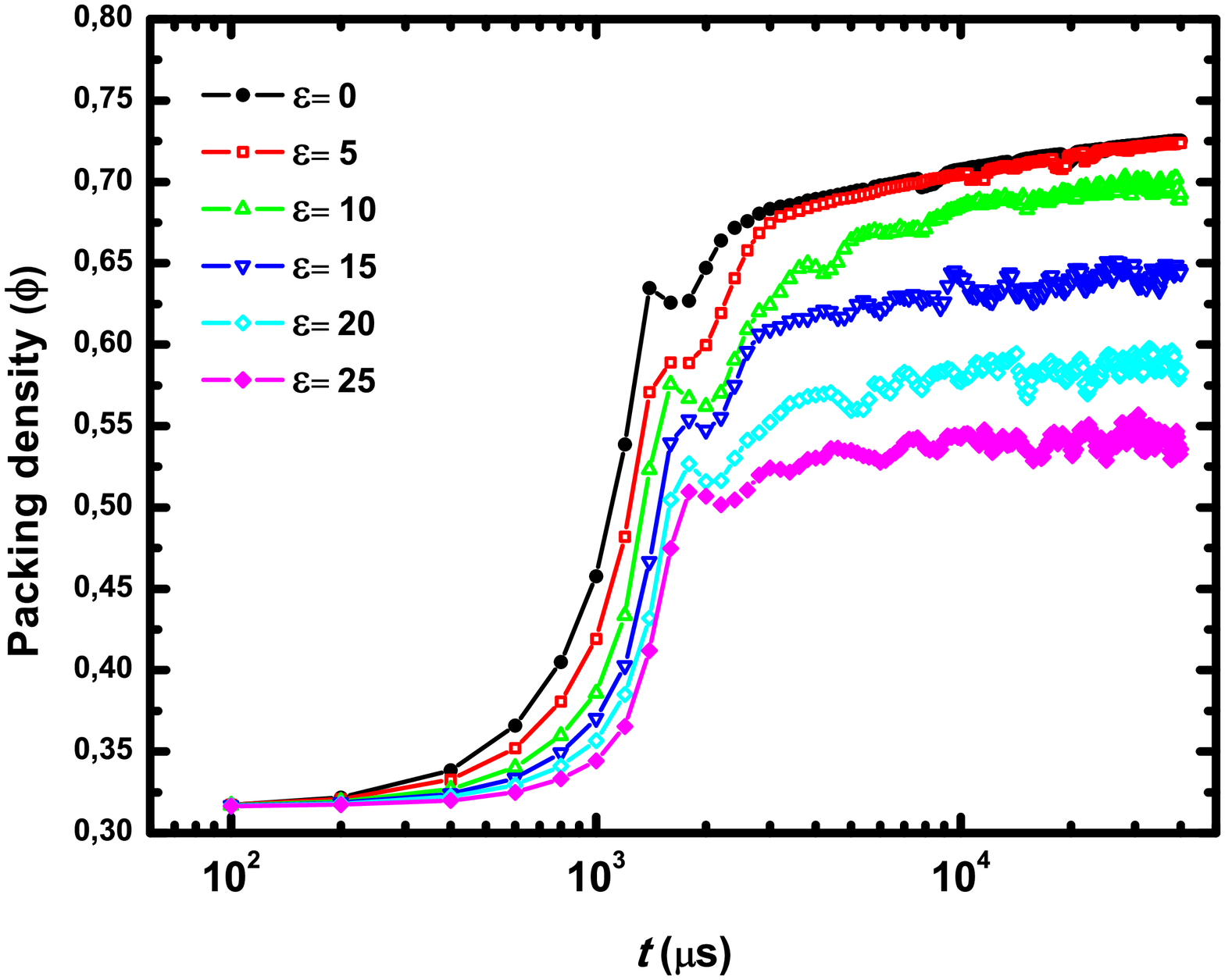}}
\end{minipage}
\caption{Plot of the packing densities $\phi$ for different probability distributions as a function of the logarithm of time. Several interaction strengths $\varepsilon$ are considered for each case.}\label{fig:03}
\end{figure*}\label{fig:03}

\section{\label{sec:r} Results and Discussion}

In this work, the particle packing processes were investigated using different size distributions and assuming different long-range interaction strengths $\varepsilon$. The initial positions, as well as the radii of $5000$ non-overlapping particles ranged from $1.0$ to $7.0 \, \mu m$ were defined inside a confining $(80 \times 80 \times 100) \, \mu m$ box by using a random number generator~\cite{recipes96}. The particles thereafter were pulled down by gravity and started to collide each other. Here, no periodic boundary conditions were assumed and, hence, the particle-wall interactions were also taken into account. The non-Gaussian distributions were represented by the L{\'e}vy and uniform distributions. The former is given by
\begin{equation} \label{eq:11}
f(r) = k\sqrt {\frac{c}{{2\pi {{(r-\mu )}^3}}}} \exp \left[ {-\frac{c}{{2(r-\mu )}}} \right],
\end{equation}
where $\mu$ is the location parameter, $c$ is the scale parameter and $k$ is a normalizing factor, which was used to normalize this distribution within the $r$ value range of interest. The latter distribution attributes equal probability of finding a given particle with radius ranging from $1.0$ to $7.0\, \mu m$ inside the box. While the Gaussian distribution is determined by the well-known form 
\begin{equation} \label{eq:12}
	f(r) = \frac{1}{{\sqrt {2\pi {\kern 1pt} {\Delta r}^2} }}\exp \left[{-\frac{{{{(r - \bar{r})}^2}}}{{2{\Delta r^2}}}}\right],
\end{equation}
being $\bar r$ the mean $r$ value and $\Delta r$ the standard deviation. Fig.~\ref{fig:01} displays the described distributions above and are defined in the range of $1.0$ to $7.0 \, \mu m$. Note that all distributions are normalized within this value range. For the L{\'e}vy distribution, two different parameter sets were considered. We termed it with parameters ($\mu=0.05\, \mu m, c=0.50\, \mu m$) as type I distribution  and with parameters $(\mu=0, c=2.0\, \mu m)$ as type II distribution. As can be seen from Fig.~\ref{fig:01}, the type I distribution generates packs that contain more small particles ($r<4\, \mu m$) than large ones ($r\geq 4\, \mu m$). Conversely, the type II distribution gives preference to larger particles rather than smaller ones during the particle radius assignment. While the Gaussian distribution is centered at $\bar r=4.0\, \mu m$ with a deviation of $\Delta r=1.5\, \mu m$. It is worth mentioning that the only constraint during the particle radius assignment was that the particles are initially non-overlapping inside the box. However, it is need to realize that the confining box imposes a certain spatial restriction over the particle size distribution. Thus, smaller particles are more easily placed inside the box than larger ones at the very beginning of the simulations. Note also that each of these distributions is nonzero at the corresponding end points.

\begin{figure*}[t]
\begin{minipage} [t]{0.49\linewidth}
\includegraphics*[scale=0.30]{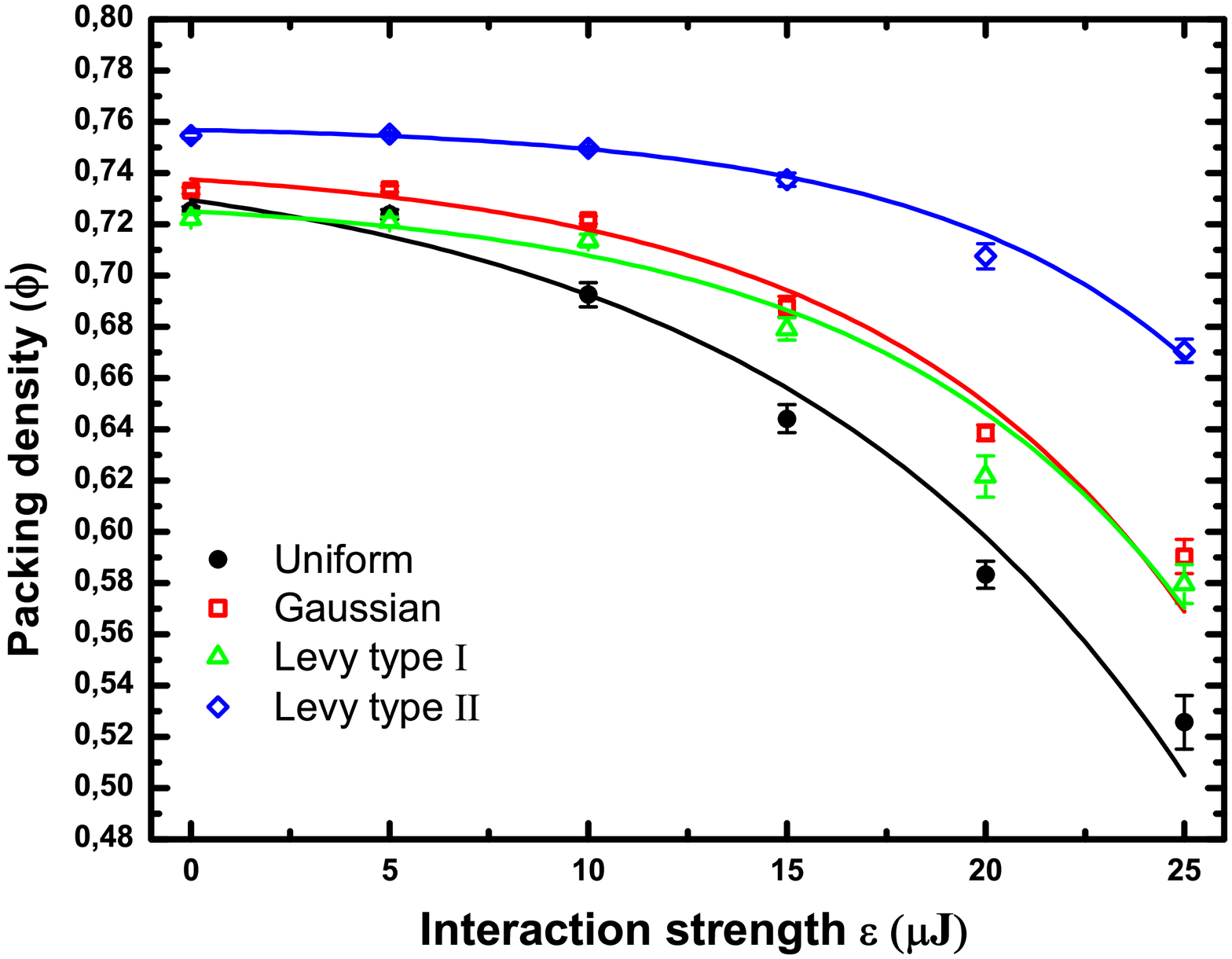}
 \caption{The ultimate $\phi$ values from Fig.~\ref{fig:03} as a function of the interaction strength $\varepsilon$. Lines are the best non-linear fits to Eq.~\ref{eq:13}. Error bars at the data points are calculated using $10$ independent runs.} \label{fig:05}
\end{minipage}\hfill
\begin{minipage}[t]{0.49\linewidth}
\includegraphics[scale=0.30]{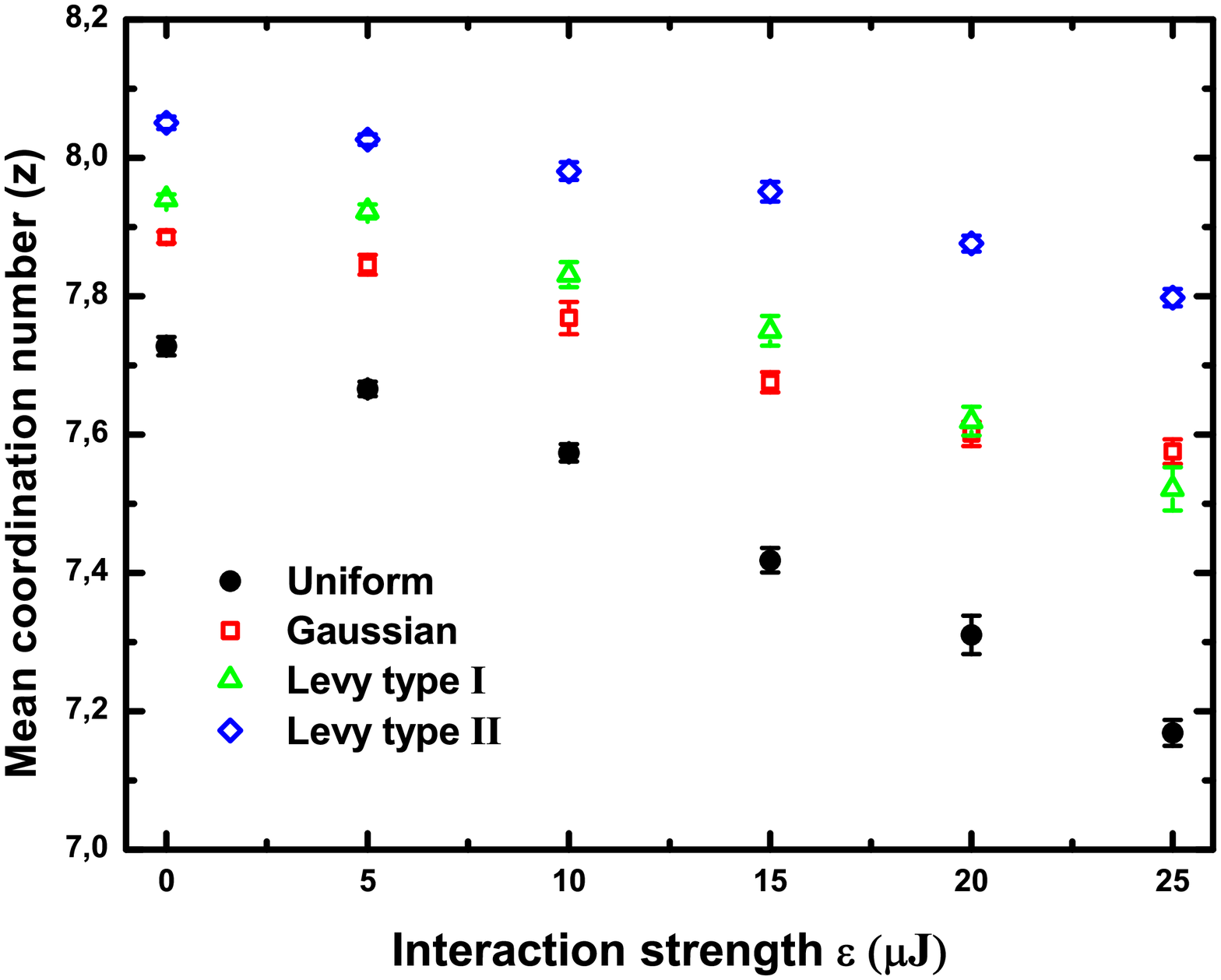}
 \caption{Plot of the  mean coordination number $z$ as a function of the interaction strengths $\varepsilon$ for all size distribution considered.} \label{fig:07}
\end{minipage}
\end{figure*}

The packing process is depicted in Fig.~\ref{fig:02} for polydispersive particles with Gaussian distribution. Snapshots at the instants $t=0.0 \, ms$ and $t=10.0 \, ms$ are shown in this figure. The parameters used in this simulation are given in Table \ref{table:01} for $\varepsilon=10.0\, \mu J$. This figure was rendered with a gray color gradient along the vertical direction (z axis) to display the different particle layers fall towards the bottom base of the box. We performed statistical calculations of the different quantities such as packing density, mean coordination number, kinetic energy, and RDF as the system evolved over time.  To determine the average value of these quantities and estimate their statistical error, we averaged over $10$ independent realizations. Furthermore, a smaller virtual box with an offset distance from each actual wall measuring $5\;\mu m$ and centered in the bulk region of the aggregate was also used to eliminate wall effects~\cite{Desmond2009} in these calculations.

In Fig.~\ref{fig:04} is shown the time evolution of the average kinetic energy per particle in the Gaussian case for three different $\varepsilon$ values. The inset shows the log-log plot of the tail of the curve after $1.0 \, ms$. Similar energy curves were found for all remaining cases. One can see that the system relaxation was already achieved around $3.0 \, ms$ for all $\varepsilon$ values considered.  Nevertheless, it is important to say that due to the long-wavelenght cooperative modes~\cite{luding94} present in the dynamics of many colliding particles as well as their own elastic properties, the total kinetic energy does not completely vanish within the times considered here. Moreover, the noise observed at large times is due mainly to persistent action of the long-range forces on the particles. From Figs.~\ref{fig:44} and ~\ref{fig:03}, we see that the system equilibration is dramatically slowed down so that no steady state has been yet reached for the cases with smaller $\varepsilon$ values ($\varepsilon=0$ to $15\, \mu J$) within the time scale considered here. On the other hand, for cases with larger $\varepsilon$ values, one can observe a typical steady state after $10.0 \, ms$ with fluctuations around an average $\phi$ value.  In particular, Fig.~\ref{fig:44} shows the time evolution of the quantity $|\phi-\phi_{u}|/\phi_{u}$ at large times in the Gaussian case for different $\varepsilon$ values. Here $\phi_{u}$ is the ultimate values obtained at $40 \, ms$ for each case. The data points were obtained by taking averages over 10 runs. The time axis is plotted on a logarithmic scale in order to better track the slow behavior of the convergence of this quantity. The inset displays the packing density curves on a linear scale for each case. As expected, fluctuations in the $\phi$ values are stronger at higher $\varepsilon$ values.

The packing densities $\phi$ for different probability distributions, considering several long-range interaction strengths $\varepsilon$, are shown as a function of time in Fig.~\ref{fig:03}. The time axis is plotted on a logarithmic scale so that more than four decades can be easily viewed. In general, it is seen that the higher $\varepsilon$ values, the stronger the fluctuations in the $\phi$ values at large times. A consequence of the persistent action of the long-range forces on the particles. In this figure, the $\phi$ values are given at short time intervals of $2.0\, \mu s$ up to $40 \, ms$. At $40 \, ms$, the ultimate $\phi$ values were obtained for each case. The initial packing densities were $0.36$ for the Gaussian case, $0.33$ for the type I case, $0.43$ for the type II case, and $0.32$ for the uniform case. The packing density minimum around $2.0\, ms$ was due to the first particles bouncing after hitting the bottom of the box. In most cases, the ultimate $\phi$ values were below $\pi/\sqrt{18}\simeq 0.74$~\cite{zamponi2008}, which corresponds to closest-packing crystal structures, namely, face-centered cubic (FCC) and hexagonal close-packed (HCP) structures. For every case, the $\phi$ value was seen to decrease with increasing interaction strength $\varepsilon$. This behavior is in accordance with the experimental results obtained by Forsyth {\it et al}~\cite{forsyth2001} for monosized particles. Moreover, one can see from Fig.~\ref{fig:03} that the packing dynamics was also quite sensitive to the distribution type used to generate the particle packs. In fact, packs with an uniform distribution displayed a broader range in the final $\phi$ values compared with other distributions. While packs with a type II distribution displayed a narrower and higher range in the final $\phi$ values. These higher $\phi$ values, particularly for the type II case, can be attributed to the existence of a great number of large particles that either are wrapped around by smaller particles or create voids that are filled by smaller ones or both. The narrow range in the final $\phi$ values found for this case can also explained by the existence of a greater number of large particles in the aggregate. The larger the particle, the larger the cutoff distance of the long-range forces ($r_{ij}=3\,(R_{i}+R_{j})$). As a consequence, the $\phi$ values, in this case, become less sensitive with a rise in $\varepsilon$. Thus, it is more difficult to compress packs with a greater number of large particles than those with small ones. On the other hand, for both the uniform (mainly due to the initial non-overlapping condition) and type I cases (where more small particles are present), a broader range in the $\phi$ values was obtained when different $\varepsilon$ values were considered.
In particular, systems simulated when $\varepsilon=0$ behave like hard-sphere ones as those studied earlier in Refs.~\cite{santos2014,Desmond2014} and yield higher densities in comparison with those ones when $\varepsilon$ is non-zero. This happens because in $\varepsilon=0$ systems one has some features of a coarse-grain packing wherein long-range forces do not take place anymore.

\begin{figure*}[!t]
 \centering
 \includegraphics[scale=0.45]{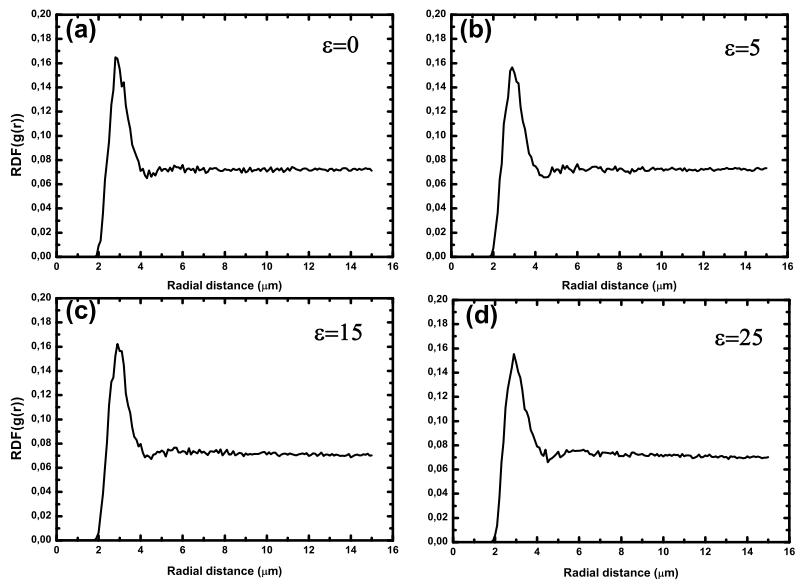}
 \caption{RDFs versus the radial distance for spheres packing structures with L{\'e}vy type I distribution for different $\varepsilon$ values. a) $\varepsilon=0$, b) $\varepsilon=5$, c) $\varepsilon=15$ and d) $\varepsilon=25\, \mu J$.} \label{fig:08}
\end{figure*}
Fig.~\ref{fig:03a} shows the time evolution of $\phi$ for the Gaussian case using several $\varepsilon$ values. Here, we obtain an ultimate density of $0,73319 \pm	0,001$ when $\varepsilon=0$ (i.e., absence of long-range forces) and $0,59 \pm 0,01$ when $\varepsilon=25\, \mu J$. Similarly, Figs.~\ref{fig:03b}, \ref{fig:03c} and \ref{fig:03d} show the time evolutions of $\phi$ for types I and II as well as the uniform case. For the type I case, we obtained $0,722 \pm 0,002$ when $\varepsilon=0$ and $0.58 \pm 0.01$ when $\varepsilon=25\, \mu J$. For the type II case, we obtained $0,754 \pm 0,001$ when $\varepsilon=0$. This is in good agreement with the experimental result of $0.746$ achieved by Ref.~\cite{sohn1968}. In fact, the systems studied in Ref.~\cite{sohn1968} were sand piles composed by polydispersive millimeter-sized particles. Such systems are good examples of coarse particle packings wherein one no longer observes long-range forces between particles. Thus, it is expected that some cases treated here when $\varepsilon=0$ yield densities closer to 0.746 obtained in certain coarse particle packings. When $\varepsilon=25\, \mu J$, we got a density of $0,67 \pm	0,005$ . For the uniform case, we obtain $0,725 \pm	0,002$ when $\varepsilon=0$ and $0,526 \pm	0,011$ when $\varepsilon=25\, \mu J$. The ultimate $\phi$ values found in Fig.~\ref{fig:03} were then plotted in Fig.~\ref{fig:05} as a function of the interaction strength $\varepsilon$ together with non-linear curve fits to the data. In each case, data trend seems to gradually decay with increasing $\varepsilon$ values. Using the following expression
\begin{equation} \label{eq:13}
	\phi  = {\phi _{\max }} - A \exp (B\, \varepsilon ),
\end{equation}
where $A$ and $B$ are fitting parameters, one can observe a good fitting of the data. Error bars at each data points were calculated using $10$ independent realizations. The fitting parameters are: $A\simeq 0.02$ and $B\simeq 0.10 \, \mu J^{-1}$ for the uniform case; $A\simeq 0.01$ and $B\simeq 0.12\, \mu J^{-1}$ for the Gaussian case; $A\simeq 0.01$ and $B\simeq 0.13\, \mu J^{-1}$ for the type I case and $A\simeq 0.002$ and $B\simeq 0.15\, \mu J^{-1}$ for the type II cases. This behavior of $\phi$ as the long-range force strength increases has been corroborated both by the experimental work~\cite{forsyth2001} and by simulations of the random close packing of disks~\cite{handrey2018}. 

In addition, we also calculated the mean coordination number $z$, it means, the mean number of neighboring particles that touch a given particle. A neighboring particle is found when the distance between two particles is equal to sum of their radii. Following the same behavior as the $\phi$ values, it can be seen that the mean coordination number $z$ also decays gradually as the $\varepsilon$ value increases for all  distributions considered. Fig.~\ref{fig:07} shows the mean coordination number $z$ as a function of $\varepsilon$. Remarkably, the $z$ value decreases more steeply as the $\varepsilon$ value increases for the uniform case. While for the type II case it decreases more smoothly with increasing $\varepsilon$ values. The $z$ values for the Gaussian and type I cases are located in an intermediate region between the two other cases mentioned.

The RDF has been widely used to characterize random structures of spherical particles~\cite{yen91,jia12}, where it can be understood as the probability of finding one particle at a given distance from the center of a reference particle. Here we define RDF as
\begin{equation}\label{eq:14}
 g(r_{i})=\dfrac{n(r_{i})}{4\pi r_{i}^{2} \delta r_{i} Z},
\end{equation}
where $n(r_{i})$ the number of particle centers within the $i$-th spherical shell of radius $r_{i}$ and thickness $\delta r_{i}$. In the above equation, $Z$ is the normalization factor given by
\begin{equation}\label{eq:15}
 Z=\sum_{i=1}^{N_{r}}\dfrac{n(r_{i})}{4\pi r_{i}^{2} \delta r_{i}},
\end{equation}
where $N_{r}$ is the total number of spherical shells considered. In the above equations, we set $\delta r_{i}=0.1 \, \mu \,m$ and $N_{r}=150$.

 Fig.~\ref{fig:08} shows the RDFs versus the radial distance for spheres packing structures with L{\'e}vy type I distribution for several $\varepsilon$ values.  Similar curves are also found for other distributions. From this figure, we can see that the general shape of the RDFs reflects the random distribution of the particles, where it is practically unchanged by the long-range interaction forces, even though they strongly influence the transient state of the packing process. It is known that random structures of particles yield RDF profiles with a single peak and plateau, whereas regular structures such as FCC and HCP ones yield RDF profiles with multiple peaks and plateaus~\cite{liu1999, handrey2018}. From Fig.~\ref{fig:08}, one sees that typical plateaus become, in general, a little more tilted as the $\varepsilon$ values increase. 

However, it is important to stress that the present results obtained through particle sedimentation mechanism may be different from those obtained by using other methods. By changing the protocol for generating such packings, one may obtain slightly different results. For instance, it is known that packings generated through collective rearrangement methods have given higher packing densities~\cite{nolan1992,nolan1993}.

\section{\label{sec:c} Conclusions}

In this study, MD simulations were performed to study the random packing process of spherical particles at micrometer scales. Both contact forces and long-range dispersive forces were taken into account in these simulations. Several size distributions were considered along with different physical quantities, including the packing density, mean coordination number, kinetic energy, and RDF. The later were computed to study the packing process so as to characterize the particle structures formed over different values of the long-range interaction strength $\varepsilon$. It was found that the packing dynamics is quite sensitive to both the distribution type and the long-range interaction strength. The simulation results showed that both the packing density $\phi$ and mean coordination number $z$  gradually decayed as the $\varepsilon$ value increased for all distributions considered. Remarkably, both $\phi$ and $z$ values decreased more steeply for the uniform distribution and more smoothly for the L{\'e}vy type II distribution as the $\varepsilon$ value increased, whereas the same values for the Gaussian and L{\'e}vy type I distributions were found to be in an intermediate region between the values of other distributions studied. The general shape of the RDFs obtained reflected the random distribution of the particles, where it was practically unchanged by the long-range interaction forces.

 Finally, long-range forces can strongly influence the packing processes, particularly by affecting important quantities as packing density and mean coordination number obtained here through particle sedimentation mechanism. That is important because its potential application to the design and fabrication of novel materials such as in sintering processes.

Future investigations will involve the study of more complex systems such as the random close packing of pairs and triplets of particles.

\section{Acknowledgements}
We wish to thank UFERSA for computational support.




\bibliographystyle{model1a-num-names}

\end{document}